# Synthesis, magnetization and magneto transport study of RECoPO (RE = La, Nd & Sm)

Anand Pal[1, 2], S.S. Mehdi[2], Mushahid Hussain[2], Bhasker Gahtori[1] and V.P.S. Awana[1,*]

[1] Quantum Phenomenon and Applications (QPA) Division, National physical Laboratory (CSIR) Dr. K.S. Krishnan Marg, New Delhi-110012, India
[2] Department of Physics, Jamia MilliaIslamia, University, New Delhi-110025, India,

We report the electrical, magneto transport and specific heat of the layered polycrystalline RECoPO (RE = La, Nd and Sm) samples. These compounds are iso-structural to recently discovered superconductor LaFeAs(O/F). Bulk polycrystalline samples are synthesized by solid state reaction route in an evacuated sealed quartz tube. All these compounds are crystallized in a tetragonal structure with space group *P*4/*nmm*. The Cobalt in these compounds is in itinerant state with its paramagnetic moment above $1.4\mu_B$ and the same orders ferromagnetically (*FM*) with saturation moment of around $0.20\mu_B$ below say 80K. Though, LaCoPO shows single paramagnetic (*PM*) to ferromagnetic (*FM*) transition near 35K, the NdCoPO and SmCoPO exhibit successive *PM-FM-AFM* transitions. Both *FM* and *AFM* transition temperatures vary with applied field. Although the itinerant ferromagnetism occurs with small saturation moment, typical anti-ferromagnetic (*AFM*) transitions ($T_{N1}$, $T_{N2}$) are observed at 69K and 14K for Nd and 57K and 45K for Sm. This *FM-AFM* transition of Co spins in NdCoPO and SmCoPO is both field and temperature dependent. The Magneto-transport of NdCoPO and SmCoPO distinctly follows their successive *PM-FM-AFM* transitions. It is clear that Sm/Nd (4f) interacts with the Co (3d) in first time synthesized Sm/NdCoPO.

*Corresponding Author: Dr. V.P.S. Awana,*
*Scientist, NPL, New Delhi-12, India*
*e-mail-awana@mail.nplindia.ernet.in:*
*Web page- www.freewebs.com/vpsawana/*

# Introduction

The quaternary equiatomic ZrCuSiAs structure type REFePnO (RE = rare-earth metal and Pn = Pnictogen like P, As, Sb, Bi) compounds were known decades before in their non-superconducting pristine form [1], and hardly could attract attention. However, the situation changed dramatically, after the discovery of superconductivity in doped oxy-pnictides compound; REFePO [2] and in particular the REFeAsO [3].Due to the renewed interest in these compounds, their synthesis and structural properties are reviewed more recently [4, 5]. Basically Superconductivity is introduced in these compounds by carrier doping in superconducting Fe-As/Fe-P layers and therefore role of 3d metal is important. Substitution of $Co^{3+}$ at $Fe^{2+}$ site effectively dope electrons in FeAs and brings about the superconductivity for around 10 to 20% of doping level at close to 15K [6-8]. On the other hand full substation of $Co^{3+}$ at $Fe^{2+}$ in REFeAsO results in non superconducting state. In fact, not only the superconducting state, but also un-doped non-superconducting oxy-pnictides show interesting behaviour and hence their study may be helpful to understand the mechanism of superconductivity in this type of compounds. The pure iso-structural, cobalt based compound i.e., RECoAsO does not show superconductivity but exhibit itinerant ferromagnetism [9,11] and interaction of $RE^{4f}$ and $Co^{3d}$ moments [12-15]. Spin fluctuations play an important role in the magnetic properties of RECoAsO compound and hence possibly in the iron-based similar structure superconductors as suggested by H. Ohta et.al.[14] The iso-stuctural LaFePO and LaNiPO are non magnetic possibly due to the strongly correlated electrons being present in their un doped form. In both these compounds 3d metal magnetic ordering is suppressed due to the reduction of their magnetic moments [16, 17]. Interestingly, in the case of LaCoPO the magnetic movement does not disappear completely because Co has odd number of electrons in 3d orbital. The LaCoPO reportedly exhibits ferromagnetic transition near 43 K in magnetization versus temperature measurement under 1kOe magnetic field. [18]

Needless to say the Co based ground state of pnictides is intriguing and complex and hence need to be probed. In this direction, though the RECoAsO is widely investigated [9-15] for both non magnetic (La) and magnetic (Nd, Sm) RE, the RECoPO on the other hand is studied only in case of non magnetic La, i.e. LaCoPO [18]. Feeling the necessity, to check if the $RE^{4f}$ and $Co^{3d}$ moments interact in RECoPO, we synthesized and studied RECoPO with both non magnetic (La) and magnetic (Nd, Sm) rare earths. We present a comparative study of structural, electrical, magnetization, magneto-transport and heat capacity of RECoPO (RE = La, Nd and Sm). To our knowledge the non magnetic (Nd, Sm) based RECoPO samples are synthesized and studied for first time. Both the magnetization and magneto-transport properties change drastically when non magnetic La is replaced by magnetic Sm and Nd.

**Experimental**

All the studied bulk polycrystalline samples of RECoPO (RE = La, Nd, and Sm) are prepared by single step solid-state reaction route via vacuum encapsulation technique [7, 11, 13]. High purity (~99.9%) rare earths (La, Nd, Sm), P, $Co_3O_4$ and Co in their stoichiometric amount are weighed, mixed and ground thoroughly using mortar and pestle under high purity Ar atmosphere in glove box. The Humidity and Oxygen content in the glove box is less than 1 ppm. The mixed powders were palletized and vacuum-sealed ($10^{-4}$ Torr) in a quartz tube. These sealed quartz ampoules were placed in box furnace and heat treated at 550°C for 12 hours, 850°C for 12 hours and then at 1150°C for 33 hours in continuum. Finally furnace is allowed to cool down to room temperature at a rate of $1^0$C/minute.

The X-ray diffraction patterns of the studied sample were taken on Rigaku X-ray diffractometer with Cu $K_\alpha$ radiation. All physical property measurements, including magnetization, magneto-transport and heat capacity are carried out on Quantum Design PPMS (Physical property measurement system) with fields up to 14 Tesla.

**Results and Discussion**

The room temperature Rietveld fitted-XRD patterns of the studied LaCoPO, NdCoPO and SmCoPO samples are shown in Fig.1. The Rietveld analysis of X-ray pattern confirmed that all the studied sample are crystallised in tetragonal phase with *P4/nmm* space group in analogy of ZrCuSiAs type structure. All the main peaks of the X-ray pattern are well indexed with a space group *P4/nmm* with only a small impurity of $RE_2O_3$ in case of NdCoPo and LaCoPO samples. The minute impurity of rare earth oxide in LaCoPO and NdCoPO and not in SmCoPO may be because La and Nd are more reactive with oxygen in comparison to Sm. The Rietveld refined parameters are in good agreement with earlier reports on similar Co based similar As based oxy-pnictides [5-13]. The lattice parameters *a*, *c* and volume are 3.966(7)Å, 8.365(4)Å and 131.62(7)Å$^3$, 3.907(2)Å, 8.177(4)Å and 124.83(9)Å$^3$ and 3.8794(5)Å, 8.0759(6)Å and 121.54(6)Å$^3$ for RE = La, Nd and Sm samples respectively. The lattice parameters are decreasing as the ionic size of the rare earth metal ion is decreasing; this shows evidence of successfully replacement of rare earth metal within same structure.

The Wyckoff positions of the atoms within fitted *P4/nmm* space group are given in Table 1.The *z* positions of RE and P are 0.1503(2) and 0.6285(4), 0.1481(2) and 0.6492(4) and 0.1451(5) and 0.6439(6) for LaCoPO, NdCoPO and SmCoPO respectively. The details of Rietveld refined structural and fitting parameters for all the studies samples are given in Table 2. It is clear from Fig.1 and Table.2 that these samples are good enough for studying their physical properties.

The measured magnetic properties of LaCoPO, NdCoPO and SmCoPO are shown in Figures 2(a), (b) and (c). Though, LaCoPO shows only single Paramagnetic (*PM*) to ferromagnetic (*FM*) transition at around 100K, the NdCoPO and SmCoPO both exhibit successive *PM-FM-AFM* transitions. The *PM* state moment of Co in these RECoPO samples is itinerant and with effective paramagnetic moment of 1.5$\mu_B$ irrespective of the RE. Further the *FM* saturation moment is also same and close to 0.20$\mu_B$. This is similar to that as observed recently for RECoAsO [10-14]. An interesting fact to be noted in these systems is that though *PM* itinerant and *FM* saturated moments for Co are same, irrespective of RE in RECoPO, the non magnetic La based compound under goes only *PM-FM* transition, the RE (Nd, Sm) ones exhibit successive *PM-FM-AFM* transitions with decrease in temperature.

To elucidate upon the temperature dependent magnetic behaviour of RECoPO, Fig. 2(a) shows the magnetization versus temperature plots for LaCoPO at various fields of 10Oe, 500Oe and 1000Oe. The compound orders ferro-magnetically (*FM*) with Curie temperature ($T_c$) near about 35K at 10Oe field. The Curie temperature is field dependent and increases significantly with increasing field, which is common for *FM* transition. The inset of fig.2 (a) depicts the isothermal *MH* plots at 10, 50, 100 and 200K with magnetic field up to 30kOe. At 5 and 25K the magnetization of the compound saturates above 2kOe but as temperature increase to 40K it shows a hump near 10kOe with a gradual change. This gradual change at 40K is due to the competition between *FM* and *PM* state, below this temperature *FM* state is dominant. At higher temperatures (100 and 200K) the *PM* state is dominant and small liner magnetization is seen. The saturation moment in *FM* state is around 0.30$\mu_B$/Co.

Figure 2(b) depicts the temperature dependent magnetization plots for NdCoPO at various magnetic fields from 10Oe to 50kOe. In low applied magnetic fields NdCoPO shows paramagnetic state from 250K to 60K and an abrupt increase in the magnetization at 60K indicating ferromagnetic (*FM*) ordering. The magnetization saturates below 60K and remains nearly the same down to 20K. Below 20K the magnetization is sharply decreased, suggesting an anti-ferromagnetic (*AFM*) transition. This situation is same for low applied fields of up to 1kOe, and change slightly at higher field of 10kOe. The successive *FM-AFM* transition is nearly disappearing at 50kOe field, at this field the magnetization is like a *PM*, but not exactly. The ordered magnetic movement under higher magnetic field is canted and it shows the paramagnet (*PM*) like behaviour. It seems the *FM-AFM* transition temperature goes down with applied field. The isothermal *MH* plots of NdCoPO are shown in inset of Fig.2 (b), which are linear at 100K (*PM* state), slightly off axis in *AFM* (2.5K) state and are *FM* like between 20-50K with small opening of the loops. Unlike as in LaCoPO, the *FM* plots are not saturated due to the contribution from trivalent magnetic Nd ions along with the ordered Co spins contribution to the total moment in the NdCoPO. It is clear from Fig. 2(b) that the magnetization of NdCoAsO exhibits successive *PM-FM-AFM* transitions with temperature.

The *MT* plots of SmCoPO in applied fields from 100Oe to 30kOe are given in Fig. 2(c). SmCoPO shows successive *PM-FM-AFM* transitions with lowering the temperature, being similar to that as in case of NdCoPO. However, there are few differences, i.e., *FM* regime in SmCoPO (65K-45K) is much narrower than as for NdCoPO (60K-20K) under an applied field of 500Oe. The saturated region of *FM* state in SmCoPO disappears fast in comparison to NdCoPO with increasing magnetic field. On the other hand, the successive *FM-AFM* transition is not disappearing in SmCoPO even up to 140kOe, which is completely disappearing at 50kOe field in NdCoPO. The transition temperature of *FM-AFM* is field dependent as we increase the field the same is shifted to lower temperature. *AFM* transition temperature is about 45K in an applied field of 100Oe and the same comes down to below 6K with increase in field to 140kOe. The *MH* plots of SmCoPO are shown in inset of Fig. 2(c), which are linear in *PM* (200K, 100K) and *AFM* states (2.5K, 5K). At intermediate temperature i.e., 20 and 50K the *MH* are *FM* like. Further at 20K, the magnetization under increasing field, first increase slowly till 20kOe, and later saturates. At this temperature compound seems to be undergoing the competing *AFM–FM* transformation. A low field the magnetic moment are not fully aligned, as field increases above 20kOe, the canted magnetic moments gets aligned in one direction and a *FM* like saturation is seen.

The temperature dependent reciprocal magnetic susceptibility of RECoPO (RE=La, Nd and Sm) is linear from 250K to 150K and curved below this temperature (not shown) in *FM* regime, but rising in *AFM* state. The Curie-Weiss fitting of the *MT* data between 250 and 150K gives the effective paramagnetic moments to be 1.4$\mu_B$, 3.58$\mu_B$ and 1.7$\mu_B$ for the LaCoPO, NdCoPO and SmCoPO respectively. These results are in agreement with similar

RECoAsO compounds [9]. Co in itinerant state may not be contributing to the magnetic moment but, Ohta et. al. [10] showed that Co adds to the total effective magnetic moment. This seems to be the situation for our RECoPO samples as well. Further, it seems that $Sm^{3+}$ ion effective moment is decreased in SmCoPO (1.7$\mu_B$), but not of $Nd^{3+}$ in NdCoAsO (3.58$\mu_B$). The total paramagnetic moment of the RECoPO in *PM* state is effectively square sum of both Co and RE moments. Needless to state that the magnetic structure by neutron scattering experiments of the first time synthesized NdCoPO and SmCoPO is very much warranted, and that could only be able to effectively resolve the complex magnetism of RE(Nd/Sm) and Co in these pnictide oxy-phosphates.

The temperature dependent resistivity of LaCoPO, NdCoPO and SmCoPO at various applied magnetic field are shown in Figures 3(a), (b) and (c) respectively. The $\rho(T)$ behaviour is metallic in the full measured (2-300K) temperature range with a slight increase in slope (improved metallic behaviour) below say 50K for all the samples. Below 50K increase in conductivity could be related to the onset of *FM* state in these compounds. The zero field metallic resistivity ratios ($\rho^{300K}/\rho^{50K}$) are 6.45, 9.85 and 8.86 for La, Nd and Sm respectively, indicating them to be good conductors. Interestingly though the zero field resistivity of LaCoPO and SmCoPO is metallic down to 2K (upper inset Figs. 3a, 3c), the NdCoPO sample exhibits an upward (negative slope) sharp step at around 18K (upper inset Figs. 3b). The zero field $\rho(T)$ behaviour of these samples (LaCoPO, NdCoPO and SmCoPO) is quite similar to that as observed for RECoAsO (RE=La, Nd, and Sm) [12-14]. The low temperature (18K) step is observed nearly at the same temperature for NdCoAsO as well and is related to the complex *AFM* ordering of Nd being mediated by interacting Nd and Co moments [8, 12]. The $RE^{4f}$ and $Co^{3d}$ moments interaction [19] is certainly taking place for Sm and Nd in RECoAsO/RECoPO. A moot question arises that why in same structure either in case of SmCoAsO [13] or presently studied SmCoPO, the interacting Sm and Co moments do not give rise to unusual ordering of Sm at higher temperatures, like as for Nd at around 16K in NdCoAsO [8, 12] or at 18K in NdCoPO. One possible reason seems could be the partial ceasing of Sm moments in paramagnetic state for SmCoAsO and SmCoPO. Also, worth noting is the fact that low temperature free ion Nd and Sm ordering appears at right temperatures of 1.6K and 4.5K for both NdCoAsO[10,12]/NdCoPO and SmCoAsO [11]/SmCoPO respectively. The free ion Nd and Sm ordering for NdCoPO and SmCoPO will be further discussed in Cp results in next sections.

As far as magneto-resistance is concerned, Figure 3(a) depicts the resistivity versus temperature under magnetic field i.e., $\rho(T)H$ plots for LaCoAsO at various applied fields of 0, 10 and 50kOe. As discussed above the compound is shown metallic resistive behaviour in the studied temperature range from 2.5K to 300K. Reasonable magneto-resistance (*MR*) appears below Currie temperature ($T_c$), reaching its maximum value at around 50K. The isothermal magneto resistivity i.e., $\rho(T)H$ of LaCoAsO at different temperatures is shown in lower inset of Fig. 3(a). The *MR*% is maximum to ~ -30% at 50K in 100kOe field with a cusp-shape, this value of *MR*% is near about 2.5 times in compare of LaCoAsO. *MR*% is at maximum at 50K and decreased at higher and lower temperature, the maximum *MR*% is -2% and -10% at 100K and 20K respectively. In Para magnetic state (200K), the *MR*% is expectedly negligible. After the compound has passed through the *FM* transition and entered into the ordered *FM* state (10, 5, 2.5K) the *MR*% is again low to less than 10%. This is natural as the *MR* is maximum, when the magnetic phase transition is taking place due to magnetic phase separation [20] and same is negligible in pure *PM* (200K) and saturated *FM* (10, 5 and 2.5K) phases [14].

The $\rho(T)H$ measurement plots for NdCoPO at various applied fields are shown in Fig. 3(b). NdCoPO exhibits similar metallic behaviour as LaCoPO, except the anomalous upturn step being discussed in previous paragraphs. The anomalous upturn $\rho(T)$ step temperature of NdCoPO decreases with increasing magnetic field, see upper inset of Fig.3 (a). Similar to that as observed earlier [8,12] for NdCoAsO. The Step like transition is precisely attributed to the ordering of $Nd^{4f}$ moments hybridizing with $Co^{3d}$ moments [8, 12]. The isothermal magneto resistivity i.e., $\rho(T)H$ with varying magnetic field at various temperatures for NdCoPO is depicted in lower inset of Fig. 3(a). The *MR*% is at maximum in the *AFM-FM* transformation region and least in both *PM* and saturated *FM* states. The same situation occurs at 2.5K, at higher field of 80kOe the maximum *MR* is nearly -12%. The unusual increase of *MR* at lower temperatures is possibly due to the $Nd^{4f}$ moments ordering and their interplay with the *FM* ordered $Co^{3d}$ spins in adjacent Co-P layers. Low temperature *MR*% increase is not seen in LaCoPO, because La is non- magnetic and hence does not influence the ordered Co spins.

Figure 3 (c) exhibits the $\rho(T)$ plots for SmCoPO at various applied fields. No, step like transition is observed in zero field resistivity of SmCoPO down to 2.5K as NdCoPO. Interestingly in SmCoPO shows the same $\rho(T)$ step like upward transition under magnetic field below 40K, see upper inset of fig.3(c). The step-like upward transition in $\rho(T)$ of NdCoPO is due to *AFM* ordering of Nd spins, the same transition in SmCoPO at applied fields indicates the possibility of Sm spins being ordered in these fields at high temperatures (~40K). The step-like transition of SmCoPO under magnetic field is similar to that as for NdCoPO without applied field. It seems $Sm^{4f}$ spins under field interact with $Co^{3d}$ spins in a similar fashion as being for $Nd^{4f}$ without field. Worth mentioning is that fact the $Nd^{4f}$ and $Co^{3d}$ interaction mediated higher $T_N$ of 18K for NdCoAsO is already reported from neutron diffraction [8, 12]. The first time synthesized similar NdCoPO need to be tested for its magnetic structure by neutron diffraction. Further the SmCoAsO and SmCoPO are still in wait for cracking of their exact magnetic structure. The $Nd^{4f}$ and $Co^{3d}$ interaction mediated higher $T_N$ of 18K for NdCoAsO reminds one of famous $PrBa_2Cu_3O_7$ high $T_c$ cuprates situation, where the Pr spins order *AFM* ($T_N$) at an unusual high temperature of 17K [21]. The unusual high $T_N$ of 17K for Pr in $PrBa_2Cu_3O_7$ and its non superconductivity are explained on the basis of extended $Pr^{4f}$ hybridisation with adjacent conducting $Cu-O_2$ planes [22].

The isothermal *MR* data of SmCoPO in various applied fields up to 100kOe at various temperatures is shown in lower inset of Fig. 3 (b). The *MR%* is lowest at low temperature *AFM* (2.5K, 5K) states. The *MR* is increasing up to -22% as the temperature decrease to 20K at applied field 10kOe and saturates at higher field. Below 50K MR% is increasing due the competition between *FM-AFM* and *FM* state dominates. As temperature goes below 20K the *AFM* state dominate and MR% decrease transitions. Also evident in are the shoulder hysteresis during increase/decrease of field at 10K.

The heat capacity of $C_P$ (*T*) of all the studied samples i.e., LaCoPO, NdCoPO and SmCoPO are shown in Figure 4. The $C_P/T$ vs T plots for their specific heats below 100K are shown in main penal. The expected Sm $T_N$ (*AFM*) is seen as a peak in $C_P(T)$ at 5.4K, and for Nd the same is below 2K in SmCoPO and NdCoPO respectively. The Nd-Co mediated Nd unusual high $T_N$ of around 18K, seen earlier in $Cp(T)$ for NdCoAsO [ 8, 12], is seen as a weak transition in $Cp(T)$ for NdCoPO as well, and is marked in Fig. 4. The $C_P/T$ vs T plots slope change first at around 80K, which roughly coincides with the *FM* ordering of Co spins. With further lowering of temperature the $C_P/T$ vs T plots slope change at around 20K and 15K with a shallow broad minimum respectively for SmCoPO and NdCoPO. These temperatures roughly coincide with the complex *AFM* ordering of $Sm^{4f}$-$Co^{3d}$ and $Nd^{4f}$-$Co^{3d}$ interplayed matrix. It seems the Sm/NdCoPO undergo three magnetic transitions i.e., $T_{c,Co}$ (~80K), the $Sm^{4f}$-$Co^{3d}$ and $Nd^{4f}$-$Co^{3d}$ interplayed *AFM* below 20K and finally $Sm^{3+}$ and $Nd^{3+}$ spins individual *AFM* at 5.4K and below 2K respectively. Interestingly the $T_{c,Co}$ (~80K) and $Sm^{4f}$-$Co^{3d}$ and $Nd^{4f}$-$Co^{3d}$ interplayed *AFM* at low temperature are not seen as distinct transition in $C_P$ (*T*) measurements. Only the distinct $Sm^{3+}$ and $Nd^{3+}$ spins individual *AFM* ordering peaks are seen in $Cp(T)$ at 5.4K and below 1.8K. The $(Nd,Sm)^{4f}$ and $Co^{3d}$ mediated *FM* and *AFM* transition in (Nd/Sm)CoPO do not occur with a distinct change in entropy and a resulting $Cp(T)$ peak but rather as a continuous change in entropy and hence no peak in heat capacity.

Summarily, we synthesized RECoPO (RE=La, Nd and Sm) in single phase and studied their complex magnetism and magneto-transport. Magnetic RE (Nd and Sm) based RECoPO are synthesized and studied for first time to our knowledge. Though, LaCoPO shows single paramagnetic (*PM*) to ferromagnetic (*FM*) transition near 35K, the NdCoPO and SmCoPO exhibit successive *PM-FM-AFM* transitions. The *FM-AFM* transition for magnetic rare earths is mediated by $(Nd,Sm)^{4f}$ and $Co^{3d}$ interacting moments in (Nd/Sm)CoPO. The transport measurements showed that all studied compounds are metallic in nature, and their magnetic ordering in particular the *AFM* is clearly reflected in magneto-transport studies. NdCoPO shows a upward step like transition in $\rho(T)$ measurement under zero field but not for SmCoPO. The transition of Co spins from ferromagnetic (*FM*) to anti-ferromagnetic (*AFM*) in (Nd/Sm)CoPO is field dependent. The heat capacity $Cp(T)$ exhibits only the low temperature RE moments *AFM* ordering and as such the Co mediated transitions (*FM-AFM* )are not seen, indicating a continuous change in entropy and hence no peak in heat capacity.

Authors thank their Director Prof. R.C. Budhani for his keen interest and encouragement for the study. Anand Pal would like to thank *CSIR* (Council of scientific and Industrial Research) for granting him senior research fellowship. Bhasker Gahtori thanks *DST* (Department of Science and Technology) for supporting him by Fast Track Fellowship Post Doctoral Program to work on doped Fe based superconductors.

Figure Captions

Figure 1: Observed and Rietveld fitted room temperature XRD patterns of LaCoPO, NdCoPO and SmCoPO

Figure 2: Magnetization ($M$) versus temperature ($T$) for (a) LaCoPO, (b) NdCoPO and (c) SmCoPO at different fields. Respective insets show the isothermal magnetization $M(H)$ plots of these compounds at various temperatures.

Figure 3: Resistivity ($\rho$) versus temperature ($T$) for (a) LaCoPO, (b) NdCoPO and (c) SmCoPO at different fields from 300 down to 2.5K. Upper insets show the expanded resistivity plots for respective samples. Lower insets show the isothermal magnetization plots ($MR$) at various temperature for respective samples.

Figure 4: Heat capacity ($C_P$) versus temperature ($T$) for LaCoPO, NdCoPO and SmCoPO at from 200 down to 2K. Inset shows the expanded $C_P/T$ versus $T$ plots for same samples.

Table 1 Wyckoff position for RECoPO (Space group: *P4/nmm*)

| Atom | Site | $x$ | $y$ | $z$ |
|------|------|-----|-----|-----|
| RE   | 2c   | 1/4 | 1/4 | $z$ |
| Co   | 2b   | 3/4 | 1/4 | 1/2 |
| P    | 2c   | 1/4 | 1/4 | $z$ |
| O    | 2a   | 3/4 | 1/4 | 0   |

Table 2 Reitveld refined parameters for RECoPO

| RE | $a$(Å) | $c$(Å) | Volume (Å$^3$) | $R_P$ | $R_{WP}$ | $\chi^2$ |
|----|--------|--------|----------------|-------|----------|----------|
| La | 3.966(7) | 8.365(4) | 131.62(7) | 2.74 | 3.84 | 2.12 |
| Nd | 3.907(2) | 8.177(4) | 124.83(9) | 2.45 | 3.16 | 1.16 |
| Sm | 3.8794(5) | 8.0759(6) | 121.54(6) | 1.99 | 2.54 | 1.20 |

Figure 1

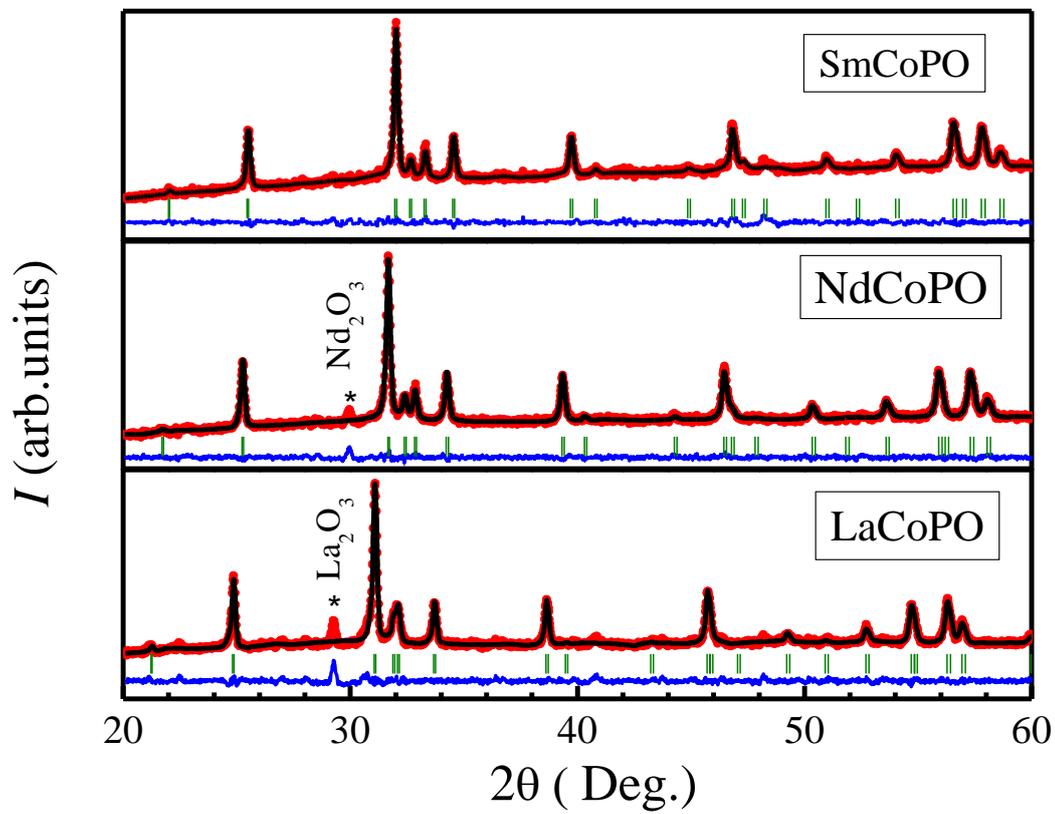

Figure 2(a)

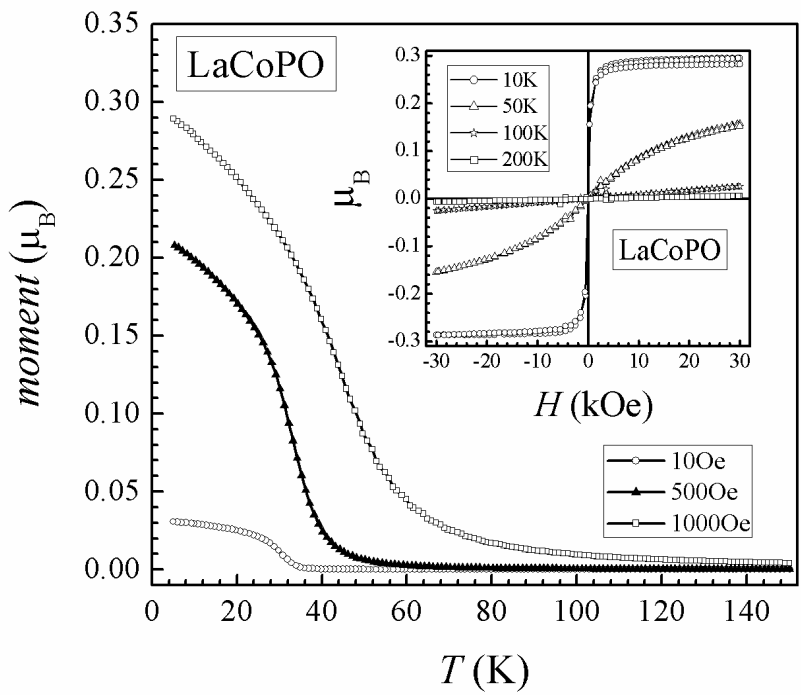

Figure 2(b)

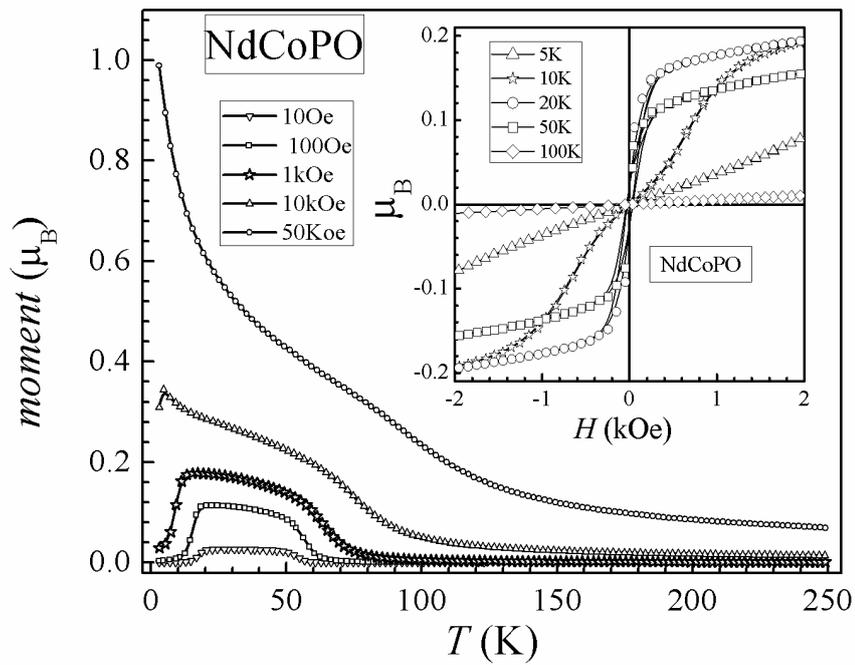

Figure 2(c)

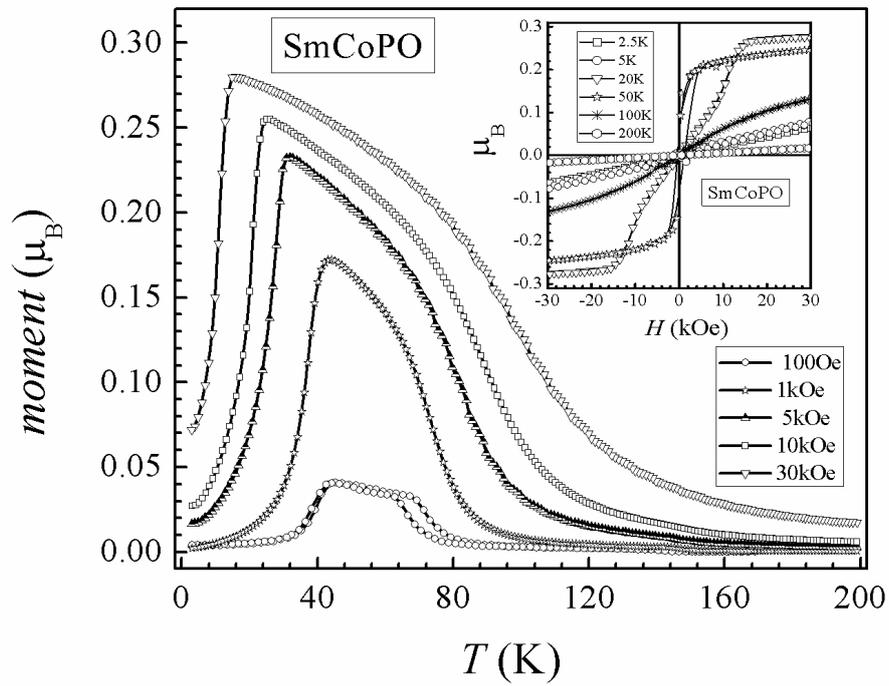

Figure 3(a)

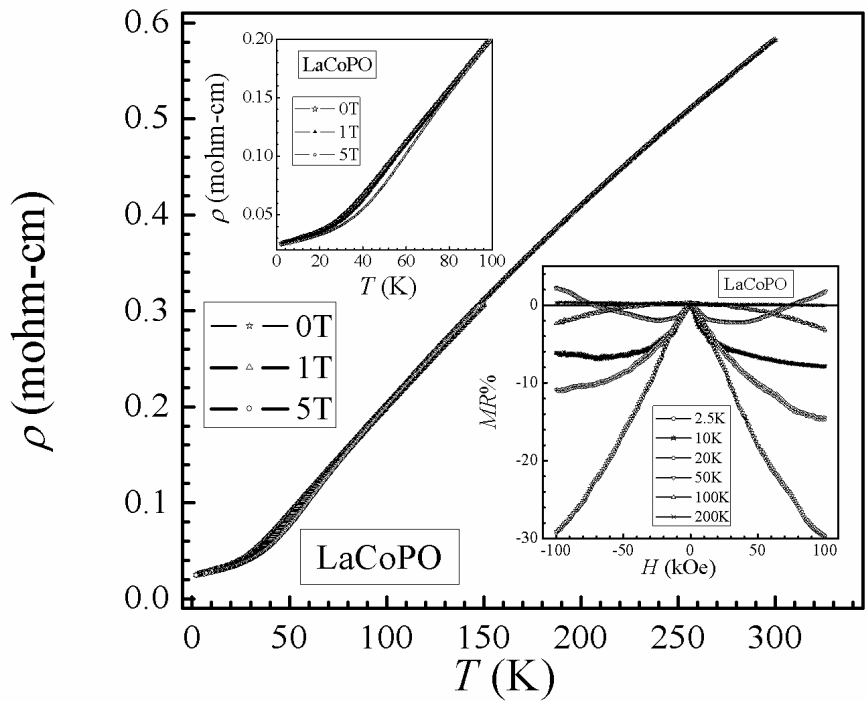

Figure 3(b)

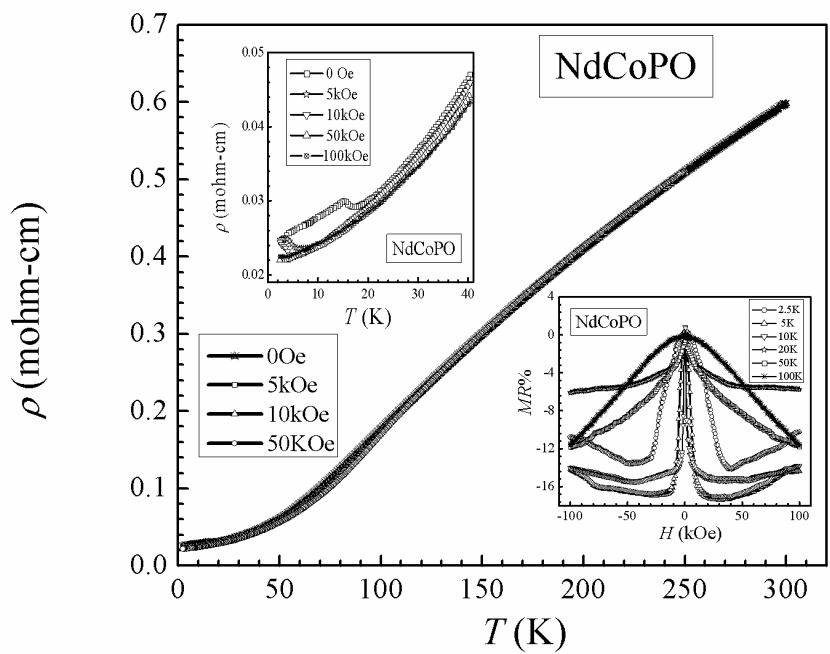

Figure 3(c)

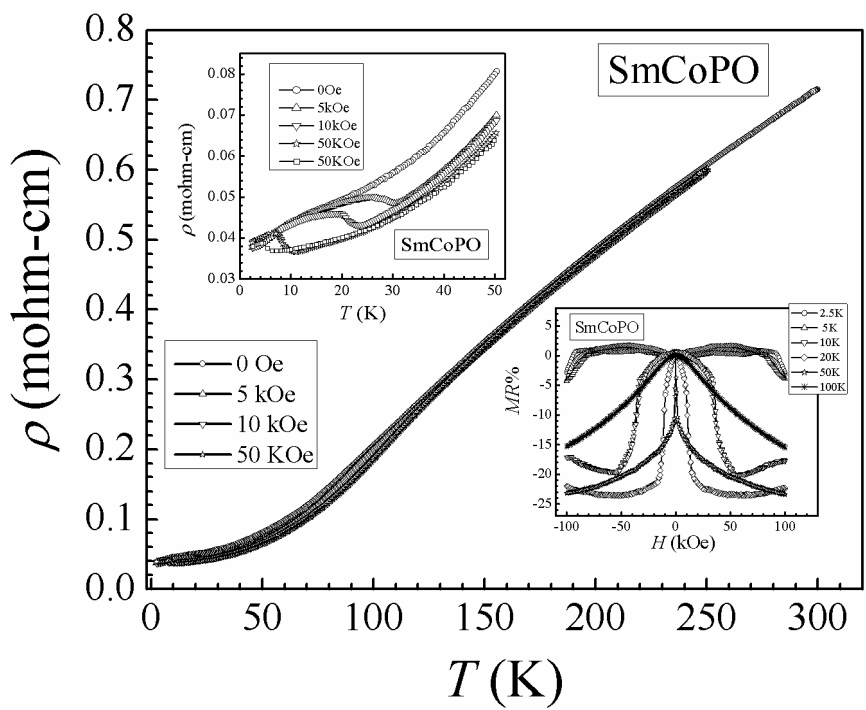

Figure 4

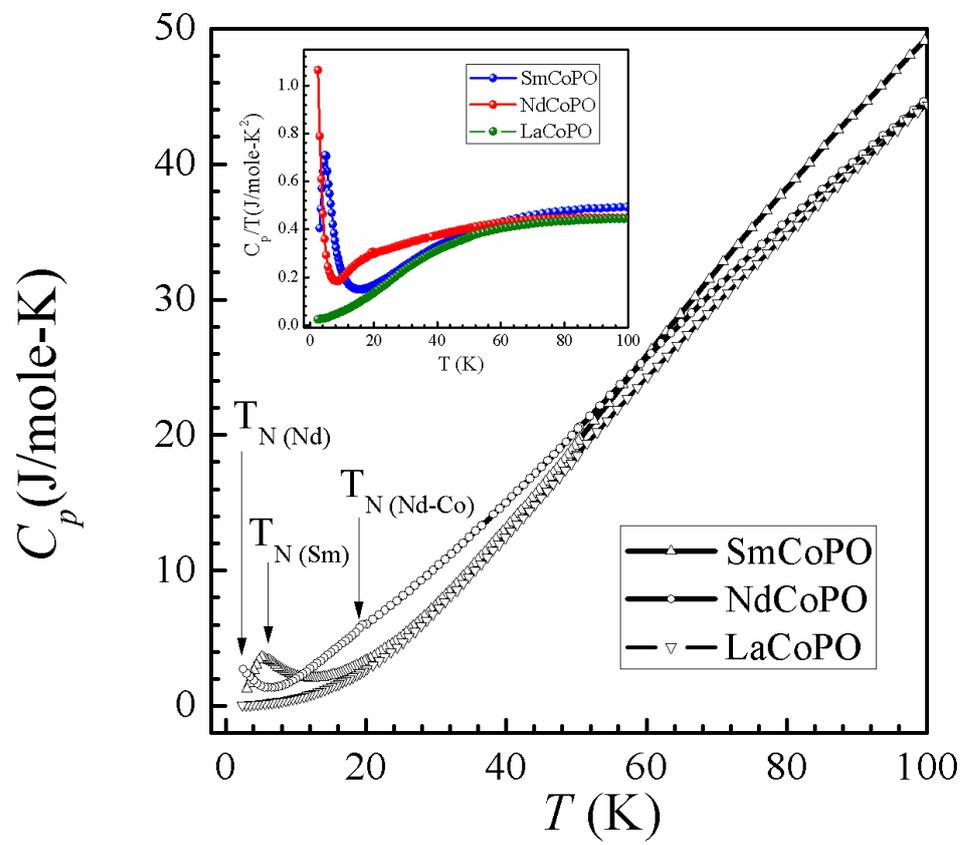